\documentclass[floatfix,preprint,showkeys,citeautoscript,showpacs,amsmath,amssymb,epsf,aip,jcp]{revtex4-1}
\usepackage{dcolumn}
\usepackage{epsfig}
\usepackage{color}
\usepackage{graphicx}
\usepackage{dcolumn}
\usepackage{bm}

\begin{document}
\title{Practical and Rigorous Reduction of the Many-Electron Quantum Mechanical Coulomb Problem to O(N$^{2/3}$) Storage}

\author{Mark R. Pederson}
\email{mark.pederson@science.doe.gov}
\affiliation{Department of Chemistry, Johns Hopkins University, Baltimore MD}                        
\date{\today}
\begin{abstract}
It is tacitly accepted that, for practical basis sets consisting of N functions, 
solution of the two-electron Coulomb problem in quantum mechanics 
requires storage of O(N$^4$) integrals in the small N limit. For localized functions, in the large N limit,
or for planewaves, due to closure, the storage can be reduced to O(N$^2$) integrals.                
Here, it is shown that the storage can be further reduced to O($N^{2/3}$) for separable basis functions.
A practical algorithm, that
uses standard one-dimensional Gaussian-quadrature sums, is demonstrated.  The resulting 
algorithm allows for the simultaneous storage, or fast reconstruction, of any two-electron coulomb integral required for a many-electron calculation, 
on each and every processor of massively parallel computers even if such processors have very limited memory and
disk space. For example, for calculations involving a basis of 9171 planewaves, the memory required to effectively store
all coulomb integrals decreases from 2.8Gbytes to less than 2.4 Mbytes.
\end{abstract}
\maketitle
 
\section{Introduction}
In this communication a workable algorithm is derived and presented that allows each processor to store all information required 
to quickly look up any two-electron integral, involving four basis functions,  needed for either density-functional or multiconfigurational
wavefunction methods. The method is demonstrated by applications of a uniform electron gas, confined to a cubic box, for electrons with 
wavevectors that are enclosed in a Fermi sphere.

Strategies for rapid calculation or efficient storage of two-electron integrals, for density-functional~\cite{HK,KS} calculations, 
or multiconfigurational active space methods~\cite{molcas,LG} continue to evolve as different mathematical techniques and different
types of computing platforms arise and as different types of basis functions are implemented for use in electronic structure
calculations. A recent comprehensive review of these efforts by Reine {\em et al}~\cite{reine} includes discussions of
least-square variational fitting methods~\cite{koster,dunlap} and
Rys polynomials~\cite{king}. Other methods such as direct methods~\cite{almlof}, analytic algebraic decompositions~\cite{nrlmol},  
tensor hypercontraction~\cite{parrish} and multipole methods~\cite{Lambrecht} are also widely used. Many of these
methods support the hypothesis that the space of two-electron integrals is smaller than naively expected. 

This paper seeks to formally prove, for separable functions used in electronic structure calculations, that the set of
information on which the N$^4$ Coulomb integrals truly depends is much smaller than expected from a permutational
analysis.
Further a practical
approach is developed and applied to the uniform electron gas.
The algorithm is                  
based upon a three-dimensional Fourier transform, a one-dimensional Laplace transform, an additional 
one-dimensional integral transform, and the use of Gaussian 
quadrature. The storage requirements needed to calculate matrix elements associated with the coulomb operator is 
reduced to O($N^{2/3}$) for either planewaves or Gaussians.

Another motivation for this work is that the
development of massively parallel methods requires one to break a problem up into many independent subtasks that can then
be performed simultaneously by a large number of computer processors~\cite{nrlmol}.  To achieve high efficiency on massively 
parallel architectures,
it is necessary to ensure that the amount of information exchanged between processors is small and that the rate of information
exchange is intrinsically faster than the computing time used by any processor. For future low-power computing platforms  
it is desireable, if not expected, for   
each processor to have a very limited amount of computer memory.  Thus, in reference to many-electron quantum mechanics or 
density functional theory~\cite{HK,PW92,PBE,MHG,DT}, 
it is appropriate to reconsider whether there are other means for reconstructing matrix elements that might be more efficient
on modern massively parallel architectures.
For such systems it would be ideal to allow
each processor to quickly reconstruct any possible coulomb integral needed for a quantum-mechanical simulation without information
transfer to or from other processors. 
\section{Derivation}
There is one important aspect of this derivation that appears to be universally correct for many, possibly all, choices of separable 
basis functions and that is definitely correct for planewave and Gaussian basis functions.
Therefore some general considerations are discussed before moving the focus of this paper to applications within planewave basis sets.
Given a set of infinitely differentiable and continuous one-dimensional functions, labeled as f$_l(x)$, it is 
possible to create  three-dimensional basis functions $g_{\bf I}({\bf r})$ according to:
\begin{equation}
g_{\bf I}({\bf r}) = f_l(x)f_m(y)f_n(z)=\prod_x f_{I_x}(x), 
\end{equation}
with ${\bf I}=(l,m,n)$.
Common examples of such basis functions include planewaves inside a box or unit cell or products of one-dimension Gaussian functions which generally
also have separable polynomial prefactors. In the former case one generally uses all possible products subject to the constraint that 
$\frac{2\pi}{L}|{\bf I}|<k_c$ and then seeks convergence by performing the calculation as a function of the cutoff wavenumber ($k_c$). 
Assuming one chooses a total of  N three-dimensional basis functions, it is then clear that there 
are approximately $N^{1/3}$ one-dimensional basis functions for each cartesian coordinate. For simplicity, but not actually required for 
this observation, the assumption is that the same one-dimensional basis sets are used for each cartesian component. So, even though 
there are $N^{2}$ pairs of three dimensional basis functions, there are only N$^{2/3}$ one dimensional products of basis functions. For
planewaves,  the complexity is further reduced to $2N^{1/3}$ since the product of a planewave is a plane wave. 
For Gaussians this number becomes $\eta N^{1/3}$, with 
$\eta$ a characteristic number of neighbors, since the product of two well separated Gaussians is identically zero. 

The matrix elements that are needed to solve the Coulomb problem in density functional theory or to determine
matrix elements required for either Hartree-Fock or Multi-Configurational calculations 
are given by

\begin{equation}
C_{\bf IJKL}=<g_{\bf I} g_{\bf J}|\frac{1}{|{\bf r-r'}|}| g_{\bf K} g_{\bf L}>=\int \int d^3r d^3r'  \frac{1}{|\bf {r-r'}|} 
g_{\bf I}({\bf r}) g_{\bf J}({\bf r})
g_{\bf K}({\bf r}) g_{\bf L}({\bf r}).  
\end{equation}
However, by using a continuous Fourier transform of $\frac{1}{|{\bf r}-{\bf r'}|}$, 
followed by a Laplace transform of $\frac{1}{p^2}$, the 
above equation can be written in quasi-separable form according to:
\begin{eqnarray}
C_{\bf IJKL}=&&<g_{\bf I} g_{\bf J}|\frac{1}{|{\bf r-r'}|}| g_{\bf K} g_{\bf L}>  \\
&&=4 \pi \int d^3p \int d^3r  \int d^3r'  \frac{e^{i{\bf p  (r-r')}}}{p^2} 
g_{\bf I}({\bf r}) g_{\bf J}({\bf r})
g_{\bf K}({\bf r'}) g_{\bf L}({\bf r'}).\\ 
&&=4 \pi \int_0^\infty d \alpha \int d^3p \int d^3r  \int  d^3r'  e^{i{\bf p  (r-r')}}e^{-\alpha p^2} 
g_{\bf I}({\bf r}) g_{\bf J}({\bf r})
g_{\bf K}({\bf r'}) g_{\bf L}({\bf r'})  \\
&&=4\pi \int_0^\infty d\alpha  
\prod_x A_x(\alpha,I_x,J_x,K_x,L_x) \\
&&=4\pi \int_0^{\alpha_c} d\alpha  
\prod_x A_x(\alpha,I_x,J_x,K_x,L_x)
  +4\pi \int_{\alpha_c}^\infty d\alpha  
\prod_x A_x(\alpha,I_x,J_x,K_x,L_x).
\end{eqnarray}
Eq.~4 follows from Eq.~3 by a continuous Fourier transform of $1/|{\bf r- r'|}$. Eq.~5 follows from Eq.~4 by a continuous
Laplace transform of $1/p^2$.  Eq.~6 follows from Eq.~5 since all functions are separable.
In the above equation, the nine-dimensional integral is reduced to a triple product. Each one of these products are
composed of  three dimensional integrals that is defined according to:
\begin{equation}
A_x(\alpha,I_x,J_x,K_x,L_x)= \int dx \int dx' \int dp_x  e^{-\alpha p_x^2}  e^{i p_x(x-x')}
f_{I_x}(x)f_{J_x}(x')
f_{K_x}(x)f_{L_x}(x').
\end{equation}
\begin{figure}[htp]\centering{ 
\includegraphics[scale=0.20]{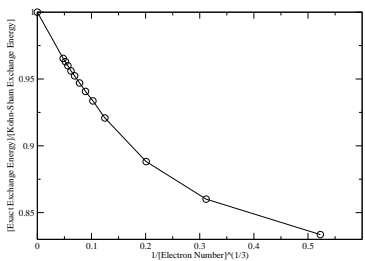} }
\caption{Ratio of the exact exchange energy to the Kohn-Sham exchange energy~\cite{KS} as a function of the
number of electrons placed inside a cubic box. The total number of electrons of each spin varies 
from M=7 (right-most point) to M=9171 (left-most point). For purposes of presentation the variable designating the number of
electrons (M) is taken to be $1/M^{1/3}$.}
\end{figure}
\begin{figure}[htp]\centering{ 
\includegraphics[scale=0.20]{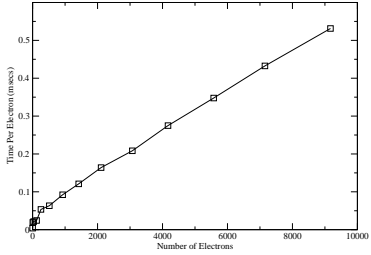} }
\caption{Time per electron, in milleseconds, on a MacBook Air, to evaluate the exchange energy as a function of the number
of electrons. The O(N$^{2/3}$) storage approach, described in this paper, shows that the time required to calculate
the  exchange energy for the uniform
electron gas increases quadratically as a function of the number of electrons. Construction of the look up table required
approximately ten minutes of MacBook Air time. The size of the look up table for a basis of 9171 planewaves requires 
less than 2.4 Mbytes of disk space. For 9171 planewaves, the memory required to simultaneously store 
all integrals, if an $O(4N^2)$) storage algorithm is used, would be approximately 2.8 Gbytes.}
\end{figure}
For either one-dimensional planewaves or Gaussians, the above three-dimensional integral can be determined, as a function of $\alpha$, 
without significant difficulty. It
is possible that for other separable functions these integrals would be difficult to calculate. 
However, since in the worst case there
are only $N^{4/3}$  of these integrals,
one can imagine calculating them only once and storing them forever.
This means that
one only needs to find an efficient numerical method for performing the Laplace integral in Eq.~6.
From this standpoint, an observation that is absolutely key to capitalizing on this quasi-separable form is 
that by integrating the above expression (Eq.~8) over $p_x$, the $\alpha$-dependent part of the, now,       
two dimensional integral, can in principal, be reduced to products of quantities with the following form:
\begin{equation}
\frac{exp(-\frac{(x-x')^2}{4 \alpha})}{\sqrt{\alpha}} = \Sigma_{n=0}^{\infty} 
a_n \frac{(x-x')^{2n}}{\alpha^{n+\frac{1}{2}}},
\end{equation}
with $a_n=(-1)^n/n!$.  Therefore, for a large enough value of $\alpha_c$, it follows that Eq.~(7) may be rewritten, to any desired precision,
according to:
\begin{equation}
C_{\bf IJKL}=
=4\pi \int_0^{\alpha_c} d\alpha  
\prod_x A_x(\alpha,I_x,J_x,K_x,L_x)
+4\pi \Sigma_{n=0}^{\infty} \Gamma_n({\bf I,J,K,L})  \int_{\alpha_c}^\infty d\alpha  
\frac{1}{\alpha^{n+\frac{3}{2}}}.
\end{equation}
In the above equation the $\Gamma_n$ are hard-to-determine constants that depend upon the functional form of separable basis sets,  
the Taylor expansion coefficients, $a_n$, in Eq.~(9), a lot 
of really complicated algebra, triple products of two-dimensional integrals associated with Eq.~(8), and the collection
of common coefficients of $1/\alpha^{n+3/2}$ arising from the occurrence of triple summations associated with each
cartesian coordinate. It would be algebraically difficult and computationally inefficient but
not impossible to calculate these numbers. 
{\bf However, for the purpose
here it is only necessary to know that the value of $\Gamma_n$ could, in principle, be found and to accept that knowledge 
about the asymptotic power law associated with the Laplace integrand provides very important information about how to numerically 
evaluate the integral which extends to infinity.}
To make further progress, the second term in
the Eq.~10 is temporarily rewritten by making the substitution $t=\frac{1}{\sqrt{\alpha}}$,  
and $dt=-\frac{1}{2}\frac{d\alpha}{\alpha^{3/2}}$. This leads to:
\begin{equation}
C_{\bf IJKL}=
4\pi \int_0^{\alpha_c} d\alpha  
\prod_x A_x(\alpha,I_x,J_x,K_x,L_x)
  +2\pi \Sigma_{n=0}^{\infty} \Gamma_n({\bf I,J,K,L})  
\int^{\frac{1}{\sqrt{\alpha_c}}}_{0}   t^{2n} dt.
\end{equation}
Now, since both definite integrals are to be evaluated over a finite interval, these integrals can be evaluated 
using Gaussian-quadrature or other one-dimensional numerical integration  meshes according to:
\begin{eqnarray}
C_{\bf IJKL}&&=
4\pi \Sigma_{i=1}^Q w_{1i}     
\prod_x A_x(\alpha_i,I_x,J_x,K_x,L_x) \\ \nonumber
+&&4\pi 
\Sigma_{i=1}^Q \frac{w_{2i} }{2}
\Sigma_{n=0}^{\infty} \Gamma_n({\bf I,J,K,L})  
\frac{\alpha_i^{3/2}}{\alpha_i^{3/2}}   t_i^{2n}.
\end{eqnarray}
In the above expressions the two sets of Gaussian-quadrature weights and points, ${w_{1i},\alpha_i}$ and
${w_{2i},t_i}$ depend only on the choice of $\alpha_c$ and methods and codes for choosing these points
are widely available and well known~\cite{vmesh,recipes}. 
A back transformation of the right-hand sum, obtained by setting $\frac{1}{\alpha_i}=t_i^2$, and defining 
$\Omega_i=\frac{1}{2}w_{2i} \alpha_i^{3/2}t_{i}^{2n}$, the integral collapses to the original recognizable form:
\begin{eqnarray}
C_{\bf IJKL}&&=
=4\pi \Sigma_{i=1}^Q w_i[0,\alpha_c]     
\prod_x A_x(\alpha_i,I_x,J_x,K_x,L_x) \\ \nonumber
+&&4\pi 
\Sigma_{i=1}^Q \Omega_i  
\Sigma_{n=0}^{\infty} \Gamma_n({\bf I,J,K,L})  
\frac{1}{\alpha_i^{n+\frac{3}{2}}}.
\end{eqnarray}
With a suitable redefinition of notation for the volume elements and the recognition that the second term includes a summation
which is  exactly equal to $\prod_x A_x(\alpha_i,I_x,J_x,K_x,L_x)$, the Laplace integral is reduced to quadratures over
products of three one-dimensional integrals (Eq.~8).  Here, it is emphasized, that Eq.~(7) could have been
immediately written in terms of numerical integrals. However the analysis followed allows one to determine how the
asymptotic form of the integrand scales so that the particular case of Gaussian quadrature 
methods, that are amenable to numerical evaluation of polynomials over
finite intervals, may be used  for performing the integrations.
As written, it has been demonstrated that one needs to store at most  
N$^{4/3}$ one dimensional integrals to reconstruct any of the N$^{4}$ integrals. Based on past
usage of quadrature methods, it is reasonable to expect that one can perform multiscale numerical
one-dimensional integration, such as the Laplace transformation here, with 
approximately 30-100 sampling points~\cite{vmesh}.                                              
\begin{eqnarray}
C_{\bf IJKL}&&
=4\pi \Sigma_{i=1}^{2Q} \Omega_i     
\prod_x A_x(\alpha_i,I_x,J_x,K_x,L_x) 
\end{eqnarray}
While the results discussed here are a factor of 2-4 away from this goal, it is likely that the number of 
sampling points can be significantly decreased by 
determining the value of $\alpha_c$ which allows for the most efficient numerical integration, 
by breaking the $\alpha$ integral (Laplace transformation)  into more than two intervals, 
and/or by using techniques similar to the variational one-dimensional exponential quadrature methods 
of Ref~\cite{vmesh}. For example, a 
quadrature mesh constructed to integrate  polynomials of $x^2$, rather than x, would be twice as efficient as 
the standard Gaussian quadratures meshes.  Except for the
clear need to exploit the $t=1/\sqrt{\alpha}$ transformation for the final interval that extends to $\infty$, 
finding the best quadrature sums are expected to depend on the form of the separable functions being employed. 
Here, for simplicity and reproducibility by others, only standard Gaussian-quadrature methods, with $\alpha_c \equiv 1$, 
are used. 

\section{Reduction of Storage to $4N^{2/3}$ for Plane Waves: Exact Exchange for the Uniform Electron Gas}
For planewaves, the product of the one-dimensional functions $f_{I_x}f_{j_x}$ reduce to a product               
of two one-dimensional planes waves which is itself a planewave. If one starts with $N^{1/3}$ one
dimensional planewaves (e.g. $f_I=exp(i2I\pi/L)$, the products will only provide $2N^{1/3}$ plane waves. Therefore the number
of one-dimensional integrals that are required is reduced to $4N^{2/3}$. 
As a simple application, the M-dependence of the exact exchange energy of an unpolarized gas of 2M electrons
in a box with finite volume (V=LxLxL) is determined in this section. As M gets very large, the 
exchange energy will converge to the Kohn-Sham value of $E_{KS}=-(3/4)(6/\pi)^{1/3} M^{4/3}/L$. It is also easy to verify based on 
scaling arguments that for any number of planewaves placed inside such a box, the exact exchange energy will 
scale a $\beta(M,\{q_{\bf k}\})$/L with $\beta$
depending on the occupations $\{q_{\bf k}\}$  as a function of wavevector and the number of electrons M placed in the box. 
Here to validate the numerics,
the standard choice of occupation numbers are taken to be unity for all planewaves enclosed in a Fermi sphere of various radii. 
The radii, or Fermi wavevector, are chosen so that there are shell closings in reciprocal space.


For a finite system, it is possible to fully occupy a Fermi sphere for a well defined cutoff wavevector if one 
chooses M=  7, 33, 123, 257, 515, 925,1419, 2109, 3071, 4169, 5575, 7153, or 9171 electrons of each spin.  In Fig.~1, the ratio of the 
exact exchange energy to the Kohn-Sham energy is presented as a function of $1/M^{1/3}$. In the large M limit, it is evident that this 
ratio converges linearly to 1. This indicates that all the integrals are being performed accurately.  In Fig.~2, the time required
per electron, as a function of the total number of electrons, is shown. For cases where each KS-orbital is identically equal to 
a planewave the time required for the calculation of the exchange (or coulomb) interaction scales as the square of the number of
electrons.  For 9171 electrons, the Hartree-Fock exchange energy can be calculated in four seconds on a MacBook Air.
In Table~I, the convergence of the
Hartree-Fock energy for M=9171 parallel spin electrons is shown as a function of Gaussian-quadrature mesh. For purposes of reproducibility,
the first mesh is determined by Q quadrature points on the interval between 0 and 1. These points, designated by $\{\alpha_i,w_{1i}\}$ in
Eq.~(12) are then transformed as described above to reduce the calculation of each exchange integral to the form shown in Eq.~14 (e.g. a total
of 2Q mesh points for the two intervals).
The results show that with standard quadrature methods, and an overly simple tesselation into only two sub-intervals, 
it is difficult to efficiently converge the energy due to sharp structure near $\alpha=0$.
However, as shown in the right-most columns, if one further breaks the first interval into sub-intervals defined by 
$[0,1/5^7],[1/5^7,1/5^6],...,[1/5,1]$ and then
uses 5-, 10-, and 15-point quadrature meshes in each of these sub intervals, convergence of the energy for M=9171 electrons is achieved.
                                                                                  
\begin{table}[th]
\begin{tabular}{|r|r|r|r|r|r|}
\hline
Mesh 1 (Q)  & Interval 1    & Total& Mesh 2 (Q)&Interval 1&Total    \\
\hline 
 90& 0.936629  &0.965018      & 8x5    & 0.936961& 0.965350 \\
105& 0.936884  &0.965272      & 8x10   & 0.937043& 0.965431 \\
120& 0.936953  &0.965341      & 8x15   & 0.937043& 0.965431  \\
150& 0.937001  &0.965389      &        &         &  \\
180& 9.937019  &0.965408      &        &         & \\
\hline
\end{tabular}
\caption{Ratio of exchange energy to Kohn-Sham exchange energy for a cube containing 2M=18342 electrons as a function of the number
of quadrature points used in Eq.~14. Mesh 1 uses Q quadrature points on an interval between 0 and 1. 
Mesh 2, which breaks interval 1 into eight sub-intervals with geometrically
varying length scales is numerically more efficient and allows for at least six-place precision. This suggests 
that the variational exponential quadrature methods, used for radial integrations in Ref.~\cite{vmesh} may
be more efficient}.
\end{table}
To summarize, this paper provides a practical and systematically improvable algorithm that reduces the storage required for the 
coulomb integrals to $O(N^{2/3})$ for the special cases of basis sets that are commonly used in electronic structure calculations. 
For the case of planewave calculations, it is only recently that researchers have begun to entertain the possibility of performing
multiconfigurational corrections using such basis sets. The results of this paper significantly lower the storage requirements needed
for either DFT, Hartree-Fock,  or multi-configurational methods based upon planewaves.  Future improvements of this method, with initial applications 
of the self-interaction correction~\cite{perzun,mrp1,mrp2} to the  uniform electron gas calculations are in progress~\cite{jianwei}. 
As compared to structurally simpler plane-wave methods, conversion of
this algorithm for use withing Gaussian-based-orbital methodologies, will require a large investment of 
programming time but are fully expected to provide
the same reduction of memory/disk requirements for reconstruction of the two-electron integrals.

\end{document}